\documentclass[a4paper,11pt]{article}
\usepackage{pos}

\renewcommand{\logo}{\relax}

\title{Towards the full NLO electro-weak corrections to Higgs boson pair production}

\author[a,b,c]{Marco Bonetti}
\author[a]{Gudrun Heinrich}
\author[d]{Stephen Jones}
\author[a]{Matthias Kerner}
\author[a]{Philipp Rendler}
\author[d,e]{Thomas Stone}
\author*[a]{Augustin Vestner}

\affiliation[a]{Karlsruhe Institute of Technology - ITP,
  Wolfgang-Gaede-Str. 1, Karlsruhe, Germany}

\affiliation[b]{Karlsruhe Institute of Technology - IAP,
  Hermann-von-Helmholtz-Platz 1, Eggenstein, Germany}

\affiliation[c]{Universität Tübingen - ITP,
Auf der Morgenstelle 14, Tübingen, Germany}

\affiliation[d]{Institute for Particle Physics Phenomenology, Durham University,
South Road, Durham, United Kingdom }

\affiliation[e]{Technical University Munich,
James-Franck-Str. 1, Garching, Germany}

\emailAdd{marco.bonetti@uni-tuebingen.de}
\emailAdd{gudrun.heinrich@kit.edu}
\emailAdd{matthias.kerner@kit.edu}
\emailAdd{stephen.jones@durham.ac.uk}
\emailAdd{philipp.rendler@kit.edu}
\emailAdd{thomas.stone@tum.de}
\emailAdd{augustin.vestner@kit.edu}

\abstract{Constraining the Higgs boson self-interaction is one of the main goals for the high luminosity phase of the LHC. A promising channel to this aim is the simultaneous production of two Higgs bosons from gluon fusion. For the interpretation of the data, precise theoretical predictions, also for differential cross sections, are needed. Following current projections this requires electroweak corrections at next-to-leading order.
We present the contributions of top-Yukawa and Higgs self-interaction-type, as well as the first steps towards inclusion of the complete NLO electroweak corrections.}

\FullConference{The European Physical Society Conference on High Energy Physics (EPS-HEP2025)\\
7-11 July 2025\\
Marseille, France\\}

\usepackage{slashed}
\usepackage{tabularx}
\usepackage{subcaption}

\usepackage{cite}

\begin{document}
\maketitle

\section{Introduction}
The simultaneous production of two Higgs bosons is the prime process to measure and constrain the Higgs boson self-interaction $\lambda$, since it already appears at leading order.
As such, it represents one of the main targets for the high luminosity phase of the LHC, where it is expected to be observed with up to 7$\sigma$ significance \cite{CMS:2025hfp}, predominantly produced from gluon fusion.
In order to extract the best possible constraints on $\lambda$, the theory uncertainties for this process should be at most at the percent level.

Previous years saw the emergence of QCD corrections to a high level of sophistication.
NLO QCD corrections including the full top-quark mass dependence are
available~\cite{Borowka:2016ehy,Baglio:2018lrj}
and have been included in calculations where higher loop orders, up to  N$^3$LO+N$^3$LL,  have been performed in the heavy-top limit~\cite{Grazzini:2018bsd,Chen:2019lzz,Chen:2019fhs,AH:2022elh}.
Analytic NLO results based on expansions in kinematic regions that basically cover the whole phase space are also available~\cite{Bagnaschi:2023rbx,Davies:2025qjr}.
Thus, the scale uncertainties are under control, while the top-mass scheme uncertainties remain an issue~\cite{Baglio:2020wgt} recently addressed in Refs.~\cite{Jaskiewicz:2024xkd,Davies:2025ghl}.

As Ref.~\cite{Bi:2023bnq} shows, the corrections originating from the electroweak (EW) sector are of a magnitude similar to the NNLO QCD corrections.
Multiple authors also calculated partial or approximate EW corrections~\cite{Muhlleitner:2022ijf,Davies:2022ram,Davies:2023npk,Heinrich:2024dnz,Davies:2025wke}, but due to the complexity of the calculation, an independent cross check of the full EW corrections is recommendable.
To this end, in the following we present the calculation performed in \cite{Heinrich:2024dnz}, where the gauge and Goldstone boson contributions have not yet been considered, and then comment on the changes required for the full calculation.

\section{Theoretical setup}
We start from a simplified model of Yukawa-type, containing only top quarks, Higgs bosons and gluons, with the unrenormalized Lagrangian
\begin{align}\label{eq:LYuk}
	\begin{split}\mathcal L_0=&-\frac14 \mathcal G_{0,\mu\nu}\mathcal G_0^{\mu\nu} + \frac{1}{2} (\partial_\mu H_0)^\dagger(\partial^\mu H_0) - \frac{m_{H,0}^2}{2}H_0^2 - \frac{g_{3,0}}{6} H_0^3 - \frac{g_{4,0}}{24} H_0^4\\
    &+ i \bar t_0 \slashed D t_0 - m_{t,0} \bar t_0t_0 - g_{t,0} H_0\bar t_0t_0 + \mathrm{constant}~,\end{split}
\end{align}
where the covariant derivative reads
\begin{equation}
	D_\mu = \partial_\mu - i g_{s,0} G_{0,\mu}^a t^a
\end{equation}
with the $SU(3)_c$ generators $t^a$. In the Standard Model, the couplings are defined as
\begin{align}
&g_{t,0} \equiv \frac{m_{t,0}}{v_0}~,& 
&g_{3,0} \equiv \frac{3m_{H,0}^2}{v_0}~,&
&g_{4,0} \equiv \frac{3m_{H,0}^2}{v_0^2}~.&
\label{eq:SMcouplings}
\end{align}
We will use these couplings and their renormalized equivalents to split the amplitude into disjunct contributions, called \emph{coupling structures}.
We define the top-quark and the Higgs boson mass in the on-shell (OS) scheme and $v=\big(\sqrt{2} G_\mu\big)^{-\frac12}$ in the $G_\mu$ scheme to determine the physical values for the couplings in eq.~\eqref{eq:SMcouplings} and calculate the corresponding two-loop amplitude.

We would like to mention that the omission of any vector or Goldstone bosons, as well as the bottom quark, discards the $SU(2)$ weak iso-spin symmetry present in the SM.
However, this is of no concern, since this model forms just a part of the total EW effects with no physical interpretation on its own.
Once gauge bosons and bottom quarks are added, the Yukawa model defined in eq.~\eqref{eq:LYuk} generates the Yukawa and Higgs self-interaction diagrams of the complete EW amplitude in unitary gauge, i.e. with decoupling Goldstone bosons.

The amplitude is written in terms of two CP-conserving form factors and tensor structures \cite{Glover:1987nx}
\begin{align}\label{eq:fullamp}
\mathcal{M}_{ab} = 
\varepsilon_{1,\mu} \varepsilon_{2,\nu} M^{\mu \nu}_{ab} =
\varepsilon_{1,\mu} \varepsilon_{2,\nu}
\delta_{ab}
\left(F_1 T_1^{\mu \nu} + F_2 T_2^{\mu \nu} \right),
\end{align}
with
\begin{align}\label{eq:T12}
T_1^{\mu \nu} &= g^{\mu \nu} - \frac{p_2^\mu p_1^\nu}{p_1 \cdot p_2}, \\
T_2^{\mu \nu} &= g^{\mu \nu} + \frac{1}{p_T^2 (p_1 \cdot p_2)} \left[ m_H^2 p_2^\mu p_1^\nu - 2 ( p_1 \cdot p_3) p_2^\mu p_3^\nu - 2 (p_2 \cdot p_3) p_3^\mu p_1^\nu + 2 (p_1 \cdot p_2) p_3^\mu p_3^\nu \right].\nonumber
\end{align}
The form factors are expanded in terms of the EW couplings from eq.~\eqref{eq:SMcouplings} as
\begin{align} 
F_i &= F_i^{(0)} + F_i^{(1)}, \\
F_i^{(0)} &= g_{s,0}^2  
\Big(
g_{3,0}\, g_{t,0}\ F^{(0)}_{i,g_3 g_t} 
+ g_{t,0}^2\ F^{(0)}_{i,g_t^2} 
\Big), \\
F_i^{(1)} &= g_{s,0}^2  
\Big(
g_{3,0}\, g_{4,0}\, g_{t,0}\  F^{(1)}_{i, g_3 g_4 g_t}
+ g_{3,0}^3\, g_{t,0}\ F^{(1)}_{i, g_3^3 g_t}
+ g_{4,0}\, g_{t,0}^2\ F^{(1)}_{i, g_4 g_t^2} \nonumber \\
& \phantom{= g_s^2 \Big(}
+ g_{3,0}^2\, g_{t,0}^2\ F^{(1)}_{i, g_3^2 g_t^2}
+ g_{3,0}\, g_{t,0}^3\ F^{(1)}_{i, g_3 g_t^3}
+ g_{t,0}^4\ F^{(1)}_{i, g_t^4} 
\Big),\label{eq:couplingStructures}
\end{align}
where $g_s=\sqrt{4\pi\alpha_s}$ is the strong coupling. $F^{(0)}_{i,j}$ are the leading-order contributions with $i\in\{1,2\}$ and $j$ the corresponding coupling structure.
Similarly, the $F^{(1)}_{i,j}$ are the next-to-leading-order, two-loop form factors, where six coupling structures are observed.
Additionally, each of the coupling structures is split into one-particle-reducible (1PR) and -irreducible (1PI) parts.
This allows fine-grained comparisons to other calculations, e.g.{} \cite{Davies:2025wke}.

\section{Calculation of the two-loop Yukawa and self-coupling correction}
Once the Feynman rules are extracted from the Lagrangian in eq.~\eqref{eq:LYuk}, \textsc{qgraf} \cite{Nogueira:1991ex} is used to generate the contributing 168 diagrams, which are distributed over the coupling structures according to Tab.~\ref{tab:numdiagrams}.
\begin{table}[b]
\centering
\begin{tabularx}{1.0\textwidth} { 
  >{\centering\arraybackslash}X
  >{\centering\arraybackslash}X
  >{\centering\arraybackslash}X
  >{\centering\arraybackslash}X 
  >{\centering\arraybackslash}X 
  >{\centering\arraybackslash}X 
  >{\centering\arraybackslash}X }
 \hline
 Type & $g_3g_4g_t$ & $g_3^3g_t$ & $g_4 g_t^2$ & $g_3^2 g_t^2$ & $g_3g_t^3$ & $g_t^4$  \\
 \hline
  1PI & 0  & 0  & 3 & 6  & 24  & 60  \\
  1PR & 12  & 6  & 1 & 6  & 24  & 26  \\
 \hline
\end{tabularx}
\caption{Number of Feynman diagrams (one-particle-irreducible and one-particle-reducible), excluding tadpole diagrams, which contribute to each of the bare coupling structures at NLO.}
\label{tab:numdiagrams}
\end{table}
These diagrams can be expressed in terms of 494 master integrals belonging to 7 integral families.
To get there, we perform an integral reduction via integration-by-parts identities.
In this specific case, we retain full symbolic dependence on $\{s, t, m_t, m_H\}$.
Apart from two insignificant exceptions, the basis is $D$-factorizing and finite; we choose integrals with up to three dots and in $\{2-2\epsilon, 4-2\epsilon, 6-2\epsilon, 8-2\epsilon\}$ dimensions.
This is facilitated by the programs \textsc{Kira} \cite{Klappert:2020nbg}, \textsc{Ratracer} \cite{Magerya:2022hvj} and \textsc{Firefly} \cite{Klappert:2020aqs} and crosschecked by a separate reduction employing \textsc{Reduze2}~\cite{vonManteuffel:2012np}.

The evaluation of the master integrals is performed using \textsc{pySecDec} \cite{Heinrich:2023til}.
Owing to \textsc{pySecDec}'s internal optimizations each integral only needs to be evaluated once per phase space point.
The bare amplitude shows spurious poles of orders $\{\epsilon^{-4}, \epsilon^{-3}, \epsilon^{-2}\}$; all cancel within the numerical precision.
The left-over single pole in $\epsilon$ cancels after renormalization.

\section{Renormalization}
The Higgs field vacuum expectation value (vev) receives corrections from higher orders in EW theory through tadpole diagrams, which requires additional steps to ensure expansion of the Higgs field around its true vacuum at the respective order.
The ways of handling the vev renormalization are known as \emph{tadpole schemes}.
We choose the Fleischer-Jegerlehner tadpole scheme \cite{Fleischer:1980ub}, since this yields gauge invariant results and reduces the number of diagrams to be included in the amplitude.
First, we perform the shift $ v_0 \rightarrow v_0 + \Delta v $, then expand the Lagrangian and, finally, require all tadpole terms to vanish.
This identifies the tadpole counterterm $\delta T$ as
\begin{equation}\label{eq:deltaT}
	\delta T= -\Delta v m_H^2=-T^H~,
\end{equation}
where $T^H$ is the sum of all tadpole diagrams with a Higgs leg.

For the parameter renormalization we define the counterterms via
\begin{align}
	\begin{split}
    H_0 &= \sqrt{Z_H}H=\sqrt{1+\delta_H}H,\\
	t_0 &= \sqrt{Z_t}t=\sqrt{1+\delta_t}t,\\
	m_{H,0}^2 &= m_H^2(1+\delta m_H^2),\\
	m_{t,0} &= m_t(1+\delta m_t),\\
	v_0 + \Delta v &= v(1+\delta_v) + \Delta v\label{eq:vev0}.
    \end{split}
\end{align}

The Higgs boson and top quark counterterms are fixed by imposing the OS scheme, so they can be extracted from self-energy insertions, while the vev counterterm $\delta_v$ is fixed in the $G_\mu$ scheme.
Since our Lagrangian in eq.~\eqref{eq:LYuk} contains none of the required particles to calculate the muon decay for the $G_\mu$ scheme, we have to rely on literature results for the finite contribution of the counterterm; we extract the expression from \cite{Biekotter:2023xle}.

\section{Results for the Yukawa and self-coupling corrections}
\begin{figure}
    \begin{subfigure}{0.45\textwidth}
        \includegraphics[width=\textwidth]{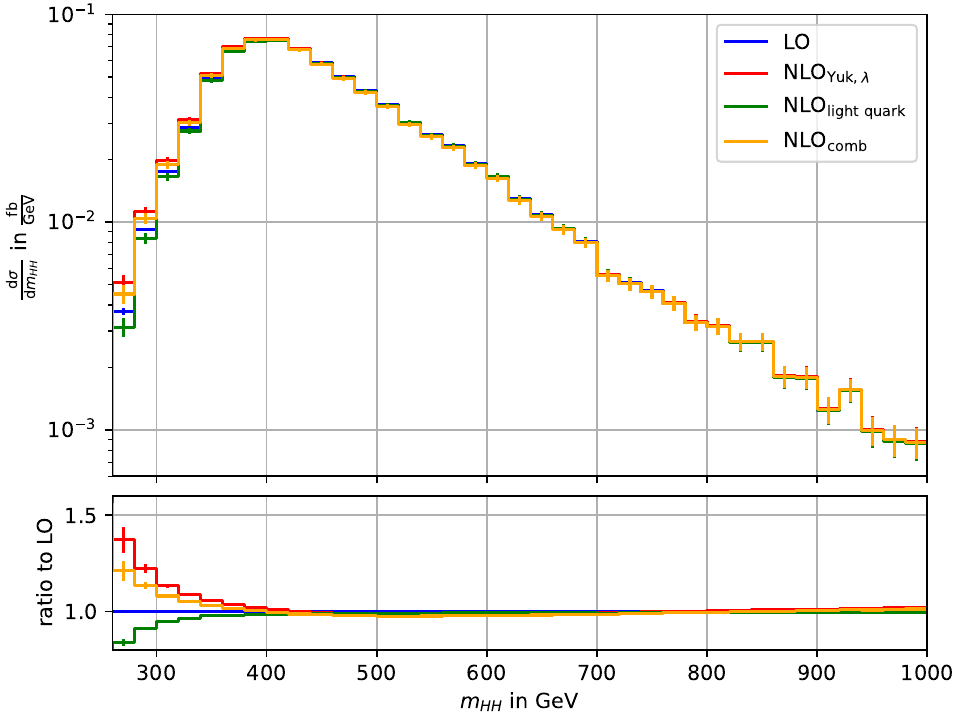}
        \caption{Differential cross section for $m_{HH}$.}
        \label{sfig:mhh}
    \end{subfigure}
    \hspace{0.1\textwidth}
    \begin{subfigure}{0.45\textwidth}
        \includegraphics[width=\textwidth]{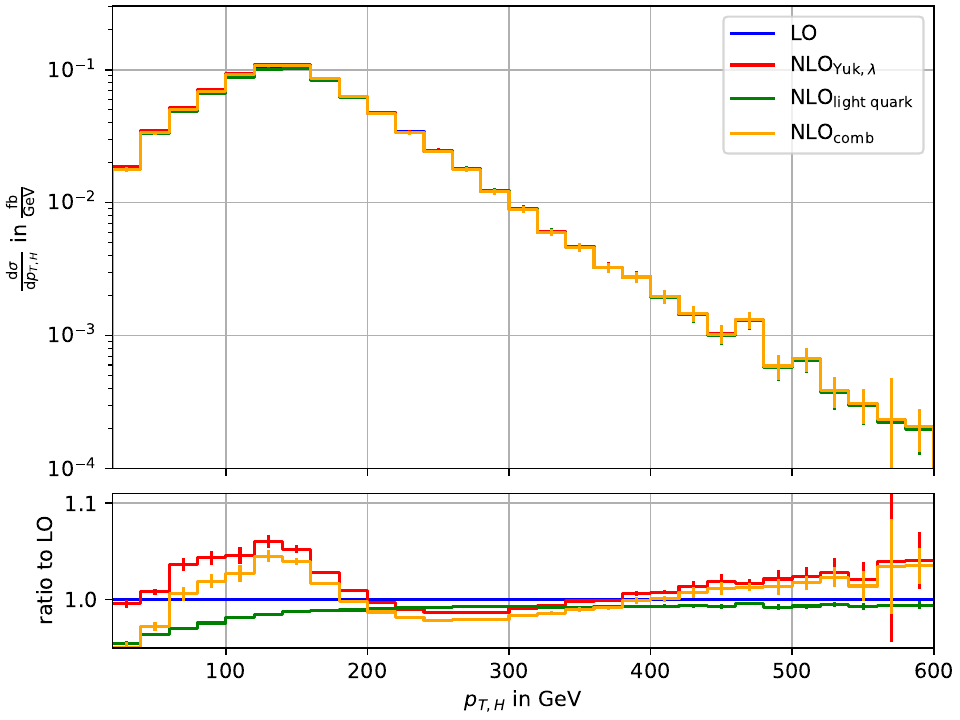}
        \caption{Differential cross section for $p_{t,H}$.}
        \label{sfig:pth}
    \end{subfigure}
    \caption{Impact of the different types of EW corrections on the $m_{HH}$ and $p_{t,H}$ distributions. The data for the light quark contribution have been calculated in \cite{Bonetti:2025vfd}.}
\end{figure}

We find an increase of 1\% in the total cross section for relevant LHC energies (13\,TeV, 13.6\,TeV, 14\,TeV), while a decrease by $\sim 4$\% has been found for the full EW corrections~\cite{Bi:2023bnq}.
In the differential cross sections, we observe a strong enhancement at low energies in the $m_{HH}$ distribution (red curve in Fig.~\ref{sfig:mhh}), even higher than for the full EW corrections, and a less pronounced but wide-spread positive contribution in the $p_{t.H}$ distribution, see Fig.~\ref{sfig:pth}.
Two-loop light-quark contributions from the authors of \cite{Bonetti:2025vfd} are also shown.
Combining the two contributions moderates the strong enhancement at low energies to a similar level as the full corrections.
In comparison, the full EW corrections in \cite{Bi:2023bnq} show a -10\% correction in the high energy tails, which also dominates the $K$-factor for the total cross section.
We conclude that, as expected, the weak bosons play a significant role in the high energy regimes of the distributions.

\section{Adjustments for the inclusion of vector bosons}
Contrary to the Yukawa model in eq.~\eqref{eq:LYuk}, the full EW sector allows CP violating tensor structures to be present in the decomposition of the scattering amplitude.
However, the new terms do not contribute at NLO as they vanish in the interference between the 2-loop and 1-loop amplitudes, the latter of which contains only CP conserving tensor structures.
Still, the Feynman rules of the EW sector contain $\gamma_5$, which must be handled in a consistent scheme.
The NLO amplitude consists of 1668 diagrams after applying several simplifications (no second generation, no leptons, diagonal CKM matrix), which yields about 1300 master integrals after reduction.
No additional counterterms beyond those computed in~\cite{Heinrich:2024dnz} are required for the complete EW calculation, as the LO coincides, however, additional diagrams do need to be considered when determining the counterterms from tadpole contributions, self-energy insertions, and vertex corrections.

\section{Conclusions}
We have presented the NLO EW corrections of top-Yukawa and Higgs-self-interaction type to Higgs boson pair production in gluon fusion and compared them to the contributions involving light quarks and vector bosons. Comparing the effects of these two partial contributions on the invariant mass of the Higgs boson pair, $m_{HH}$, and on the transverse momentum distribution $p_{t,H}$, we find that these contributions partly compensate each other, in particular at low $m_{HH}$ values. The enhancement at large values of $p_{t,H}$ is expected to be compensated by the contributions containing vector bosons and the complete doublet of third generation quarks, which are currently under construction.

\section*{Acknowledgements}
This research was supported by the Deutsche Forschungsgemeinschaft (DFG, German Research Foundation) under grant 396021762 - TRR 257, and by the UK Science and Technology Facilities Council under contract ST/T001011/1. 
SJ is supported by a Royal Society University Research Fellowship (URF/R1/201268, URF/R/251034).

\bibliographystyle{JHEP}
\bibliography{lit.bib}

\end{document}